\documentclass[conference]{IEEEtran}
\IEEEoverridecommandlockouts
\usepackage{cite}
\usepackage{amsmath,amssymb,amsfonts}
\usepackage{algorithmic}
\usepackage{graphicx}
\usepackage{textcomp}
\usepackage{xcolor}

\def\BibTeX{{\rm B\kern-.05em{\sc i\kern-.025em b}\kern-.08em
    T\kern-.1667em\lower.7ex\hbox{E}\kern-.125emX}}
\begin{document}

\title{T-Cash: Transferable Fiat Backed Coins\\
\thanks{This work was supported, in part, by Science Foundation Ireland grant 13/RC/2094 and co-funded under the European Regional Development Fund through the Southern \& Eastern Regional Operational Programme  to Lero - the Irish Software Research Centre (www.lero.ie)}
}

\author{\IEEEauthorblockN{Hitesh Tewari}
\IEEEauthorblockA{\textit{School of Computer Science and Statistics} \\
\textit{Trinity College Dublin}, Ireland\\
htewari@cs.tcd.ie}
}

\maketitle

\begin{abstract}
Numerous electronic cash schemes have been proposed over the years - however none have been embraced by financial institutions as an alternative to fiat currency. David Chaum's `ecash' scheme was the closest to something that mimicked a modern day currency system, with the important property that it provided anonymity for users when purchasing coins from a bank, and subsequently spending them at a merchant premises. However it lacked a crucial element present in current fiat-based systems - the ability to continuously spend or transfer coins. Bitcoin reignited the interest in cryptocurrencies in the last decade but is now seen as more of an asset store as opposed to a financial instrument. One interesting thing that has come out of the Bitcoin system is blockchains and the associated distributed consensus protocols. In this paper we propose a transferable electronic cash scheme using blockchain technology which allows users to continuously reuse coins within the system.
\end{abstract}


\section{Introduction}
In today's increasingly digital world we are surrounded by electronic payment systems in everyday life. People regularly use debit and credit cards to make payments at point-of-sale terminals, and on the Internet with e-commerce retailers. More and more people are also starting to make use of electronic funds transfer mechanisms, as banks try to reduce their costs and eliminate the use of cheques. Near field communications systems combined with mobile technology promises to bring more convenient and quicker ways for users to pay and conduct peer-to-peer (P2P) funds transfer. However, for all the technological advances over the past three decades cash still remains king. According to a recent report by the San Francisco Federal Reserve Bank two-thirds of all transactions below \$10, and nearly half of all transactions below \$25 are in cash \cite{FRBSF:bennett}.

It is clear that what is required is an electronic version of cash that is anonymous and with the ability to transfer coins between users without the oversight of a central authority. A scheme that was proposed by David Chaum \cite{Crypto82:chaum} more than thirty years ago fulfilled some of these requirements - however it failed to take off as a real alternative to physical cash. One of the reasons for its failure was that the scheme lacked an important property that most users associate with fiat currency - \textit{transferability of coins}. Within the scheme it was possible to anonymously withdraw coins from a bank, and at a later date spend them at a merchant. In order for the coins to be accepted by the merchant they had to be deposited with the bank for verification. However, once deposited the merchant was not able to reuse the coins i.e. transfer them to another user as change, reuse them to buy more goods etc.

Since then the arrival of Bitcoin \cite{URL:satoshi} has changed the cryptocurrency landscape beyond imagination. However, as we have seen in the past few years there is huge amounts of volatility with the value of Bitcoin. One of the contributing factors for this volatility is that Bitcoin and other Altcoins \cite{URL:hileman} have an upper limit on the number of coins that can be minted. This has led to the ``hoarding" of coins, in the anticipation that the future value of the coins will be greater than their current purchase price.

In this paper, we introduce a fiat backed, transferable electronic cash scheme with limited identification properties. Our system uses a \textit{blind signature} protocol such as the ones proposed in \cite{Crypto82:chaum} \cite{Crypto00:abe} to hide the coin transaction parameters (e.g. the serial number of the coin from the bank). We make use of blockchain technology to collectively verify the authenticity of transactions to prevent double-spending and for the transfer coins within the system, without the need to employ the services of a trusted third party (TTP).

\section{Related Work}
In this section we provide the reader with a brief background on the blind signature protocol, and some aspects of distributed ledger technologies that we make use of in the proposed protocol.

\subsection{Blind Signature Protocol}
The first electronic cash protocol that gained widespread notice by the cryptographic community was proposed by David Chaum \cite{Crypto88:chaum}. The scheme was essentially an on-line software solution whereby a buyer could spend ecash coins with any merchant participating in the system. By examining the coins (alone) neither the issuer nor the merchant were able to determine the identity of the customer. The protocol was designed such that the issuer was not able to detect the serial numbers of coins that it issued to users of the system (at the time of issue), even if it colluded with other participants in the system.

The scheme uses a blind signature protocol which allows a user (Alice) to mint a coin and hide the serial number of the coin using a blinding factor. Alice forwards the unsigned blinded coin to the bank. As long as the coin satisfies certain criteria, the bank signs the coin with its private coin signing key, without knowledge of the serial number. This feature allows for \textit{anonymous cash}. On receiving the signed coin back from the bank, Alice removes the blinding factor, and uses the coin to pay for goods at a merchant (Bob) participating in the system. The blinding factor is a random number used to obfuscate the serial number of the electronic coin from the bank. On receipt of the coin, Bob immediately forwards it to the bank for verification. The bank maintains an ever-growing database of the serial numbers of all coins that have been spent in the system and is thus able to detect double-spending.\newline

Mathematically the blind signature scheme comprises of the following steps:

\begin{itemize}
\item A set $\mathcal{M}$ which is the encrypted and non-encrypted message data and two functions $pk, sk : \mathcal{M} \to \mathcal{M}$ which are an asymmetric encryption/decryption pair:

\begin{itemize}
\item $sk(pk(m))= m$ and $pk(sk(m)) = m$ for all $m$ in $\mathcal{M}$;
\item $pk(m)$ is relatively easy to compute for all $m$ in $\mathcal{M}$;
\item $pk$'s effect on an $m$ in $\mathcal{M}$ is extremely difficult to undo (invert) without knowing $sk$, i.e. if we are given an $e$ in $\mathcal{M}$ which is the image of some $m$ in $\mathcal{M}$ under $pk$ ($e=pk(m)$) and where this $m$ is unknown then it is extremely difficult to compute $m$ without knowing $sk$.
\end{itemize}

\item There is a binary `product' $\cdot : \mathcal{M} \times \mathcal{M} \to \mathcal{M}$ on $\mathcal{M}$ the set of encrypted and decrypted message data which forms a group $\langle \mathcal{M}, \cdot \rangle$ with unit 1 (identity element) over $\mathcal{M}$.

\item The decryption function $sk: \mathcal{M} \to \mathcal{M}$ distributes over the binary product $\cdot$ on $\mathcal{M}$ i.e. $sk : \langle \mathcal{M}, \cdot \rangle \to \langle \mathcal{M}, \cdot \rangle$ is a group homomorphism.

\end{itemize}

Let ${m}$ be the coin's serial number, $(pk, sk)$ be the bank's asymmetric encryption/decryption key pair, ${r}$ is a random element chosen from the group $\mathcal{M}$. The sender encrypts ${r}$ using the bank's public key forming the blinding factor $(pk(r))$, and computes the product of the serial number with this blinding factor to form the blinded coin serial number:
\[m' = m \cdot pk(r)\]

The bank in turn signs the blinded serial number with its private key:
\[s' = {sk(m')}\]

Returns the coin to the user who removes blinding factor:
\[s = s' \cdot r^{-1}\]

The user now has a coin signed with the bank's private key:
\[
\begin{array}{cl} 
s & = {sk}(m') \cdot r ^{-1} \\
   & = {sk}(m \cdot {pk} (r)) \cdot r^{-1} \\
   & = ({sk}(m) \cdot {sk} ({pk}(r))) \cdot r^{-1} \\
   & = ({sk}(m) \cdot r) \cdot r^{-1} \\
   & = {sk}(m) \cdot (r \cdot r^{-1}) \\
   & = {sk}(m) \cdot 1 \\
   & = {sk}(m) \\
\end{array}
\]

\subsection{Bitcoin and Blockchains}
Bitcoin is a decentralized, pseudo-anonymous electronic cash scheme \cite{URL:satoshi}. The Bitcoin protocol is decentralized in the sense that the participants collectively verify all of the transactions in the network. The security of Bitcoin is based around the assumptions that a majority of the nodes in the network are honest to protect against sybil attacks \cite{LNCS:douceur}, and that the computational resources required to thwart the proof-of-work (PoW) algorithm will be greater than 50\%.

All transactions on a Bitcoin network are stored in a public ledger known as the \textit{blockchain}. The blockchain is an immutable, timestamped public ledger of all transactions that have ever been conducted on the Bitcoin network. A block in the blockchain consists of a block header and a number of associated individual transactions which are readable by all parties within the Bitcoin network.

The first block in the chain is known as the ``genesis block", followed by blocks that have been created by \textit{miners}. Miners in the Bitcoin network are nodes that compete to try to be the first to add the next valid block into the blockchain, so that they can earn some bitcoins and or transaction fees. Valid transactions are irreversibly locked into the blockchain using the PoW \cite{URL:black} algorithm by the miners, who work for a reward for solving the next PoW problem.

Each new block contains one or more new transactions that have been received by the miner within a specified time interval (e.g. every ten minutes). These are repeatedly hashed in pairs to form a Merkle tree \cite{IEEE:merkle}. The root of the Merkle tree along with the hash of the previous block is stored in the block header thereby chaining all the blocks together. This ensures that a transaction cannot be changed without modifying the block that records it and all previous blocks. This property of the blockchain makes double spending of bitcoins difficult.

Alternative consensus mechanisms to \textit{forge} the next block based on proof-of-stake (PoS) \cite{ETHEREUM:buterin}, directed acyclic graphs (DAGs) etc. have been gaining attention in the cryptographic community in recent years.

\section{System Overview}
We combine a blind signature protocol, along with the discrete logarithm problem (DLP) and a distributed ledger to allow for the transfer of coins from one user to another, without the need for a TTP to approve the transactions. The design allows for multiple independent banks to operate and mint coins in the system, such that it mirrors current global financial networks. 

Briefly a user creates a coin transaction with a number of parameters, the hash of which he blinds using a secret random quantity. He then presents this blinded hash to his bank along with his account details for the bank sign i.e. mint a coin. The bank deducts the user's account for the value of the coin, signs the hash using its secret signing key for the specified value, and returns the signature. Each bank in the system will deposit the corresponding amount of fiat currency for the coins that it mints into a common escrow account monitored by the participating banks in the system\footnote{When a coin is deposited back into a user's bank account, their account will be credited from the escrow account - thereby allowing users associated with different banks to accept coins from each other}.

The user is able to remove the blinding factor and is left with a signed hash of the coin parameters. The user then forms the coin which includes the coin parameters in plaintext along with the bank's signature. The coin is broadcast to the P2P network, and once its authenticity has been verified by the nodes in the network it is queued up by the mining nodes to be included in the next block.

In order for a user to transact with the coin they must reveal the secret DLP component to the payee who will locate the coin on the blockchain and verify that it belongs to the payer. The payee will then create a new coin transaction which will consist of a number of parameters including the revealed DLP secret, and a hash of the previous transaction, and will ask the payer to blindly sign a hash of the new transaction parameters in order to complete the coin transfer. The payee will then broadcast the new coin to the P2P network for verification and inclusion in the blockchain.

By including the secret DLP parameter and the hash of the previous transaction we are able to link various instances of the coin on the blockchain, and ensure that the coin is transferred correctly from one user to another. In essence our blockchain consists of a \textit{series of linked transactions} associated with valid coins in the network.

In the subsequent sections we provide the reader with details of our coin structure, the coin minting process, and the steps involved in transferring a coin from a payer to a payee.

\subsection{Coin Structure}\label{Overview}
A coin $(c)$ in our system is a tuple which comprises of the coin transaction parameters $(T_i)$ in plaintext and the associated blind signature, where the blind signature is of the general form $sk_i(H(T_i))$ - where $(H(T_i))$ is the hash of the transaction.\footnote{In the case of first transaction the blind signature by the minting bank is of the form $sk_{val}(H(T_i))$ - where $val$ represents the fiat value of the coin}. Each subsequent coin transaction entry on the blockchain reveals a secret quantity associated with the previous transaction (proving ownership of the coin to the intended recipient), and also ties in a new coin secret and public coin transfer key of the new owner - thereby transferring ownership of the coin. The first three components of the coin's transaction parameters $(sn, val, bank)$ are used to uniquely index the coin within the global blockchain.

\begin{center}$c = \langle T_i, sk_i(H(T_i))\rangle$\end{center} \begin{center}$T_i = (sn, val, bank, DLP_i, pk_i, x_{i-1}, H(T_{i-1}))$ \end{center} \begin{center}$DLP_i = ( p_i, \alpha_i, \beta_i)$, $pk_i = (e_i, n_i)$\end{center}

\begin{itemize}
\item A serial number $(sn)$ for the coin.

\item The fiat value $(val)$ associated with the coin.

\item The identifying financial institution $(bank)$ that minted the coin.

\item A discrete logarithm $\alpha_{i}^{x_{i}} \equiv \beta_{i} (mod~{p_{i}})$ which we denote as $DLP_i$ where:

\begin{itemize}
\item $p_i$ is a large prime which allows the formation of the multiplicative group~$\mathbb{Z}_{p_{i}}^{*}$. $p_{i}$ must be chosen in such a way that the multiplicative group $\mathbb{Z}_{p_{i}}^{*}$ contains a large subgroup $G$ of prime cardinality (as the cardinality of $\mathbb{Z}_{p_{i}}^{*}$ is $p_{i} - 1$ which is not prime). Doing this prevents Pholig-Hellman attack on the DLP \cite{IEEE:pholig}.

\item Elements $\alpha_{i}$ and $x_{i}$ from the subgroup $G$ with prime cardinality where $\alpha_{i}$ is chosen to be a primitive element of the subgroup (which will exist because $G$ has been chosen to have prime cardinality).

\item $x_i$ is the secret quantity which must only be revealed when transferring the coin to another user.
\end{itemize}

\item $x_{i-1}$ is the DLP secret associated with the previous transaction.

\item The public key component $(pk_i)$ of a RSA \textit{coin transfer} key pair $(pk_{i}, sk_{i})$.

\begin{itemize}
\item This is an ephemeral key pair which is generated by the current coin owner on a per transaction basis. This allows the owner of a coin to create a \textit{delegated signature} when transferring the coin.

\item The public key $(pk_i)$ is embedded into the coin transaction entry $(T_i)$ and is locked into the transaction by the blind signing of the previous coin's owner or the bank when the coin is minted. During subsequent coin transfers each owner creates their own delegated signature on the transaction parameters, thereby eliminating the need to contact a TTP.
\end{itemize}

\item The hash of the previous transaction data $H(T_{i-1})$.

\end{itemize}

The value $x_{i-1}$ is the DLP secret of the previous coin transaction $(T_{i-1})$ and is only revealed when proving ownership of the coin. In the case of $T_{i=1}$ the value of $x_{i-1}$ is \textit{null} as this is the first coin transaction. In addition, we chain all of the coin transactions on the blockchain together by including the hash of the previous transaction $H(T_{i-1})$ in the next transaction $(T_{i})$. Again for transaction $T_{i=1}$ the value of $H(T_{i-1})$ is \textit{null} as it is the first coin transaction.

\subsection{Minting a Coin}
\begin{figure}[h]
\centering
\includegraphics[width=3.5in]{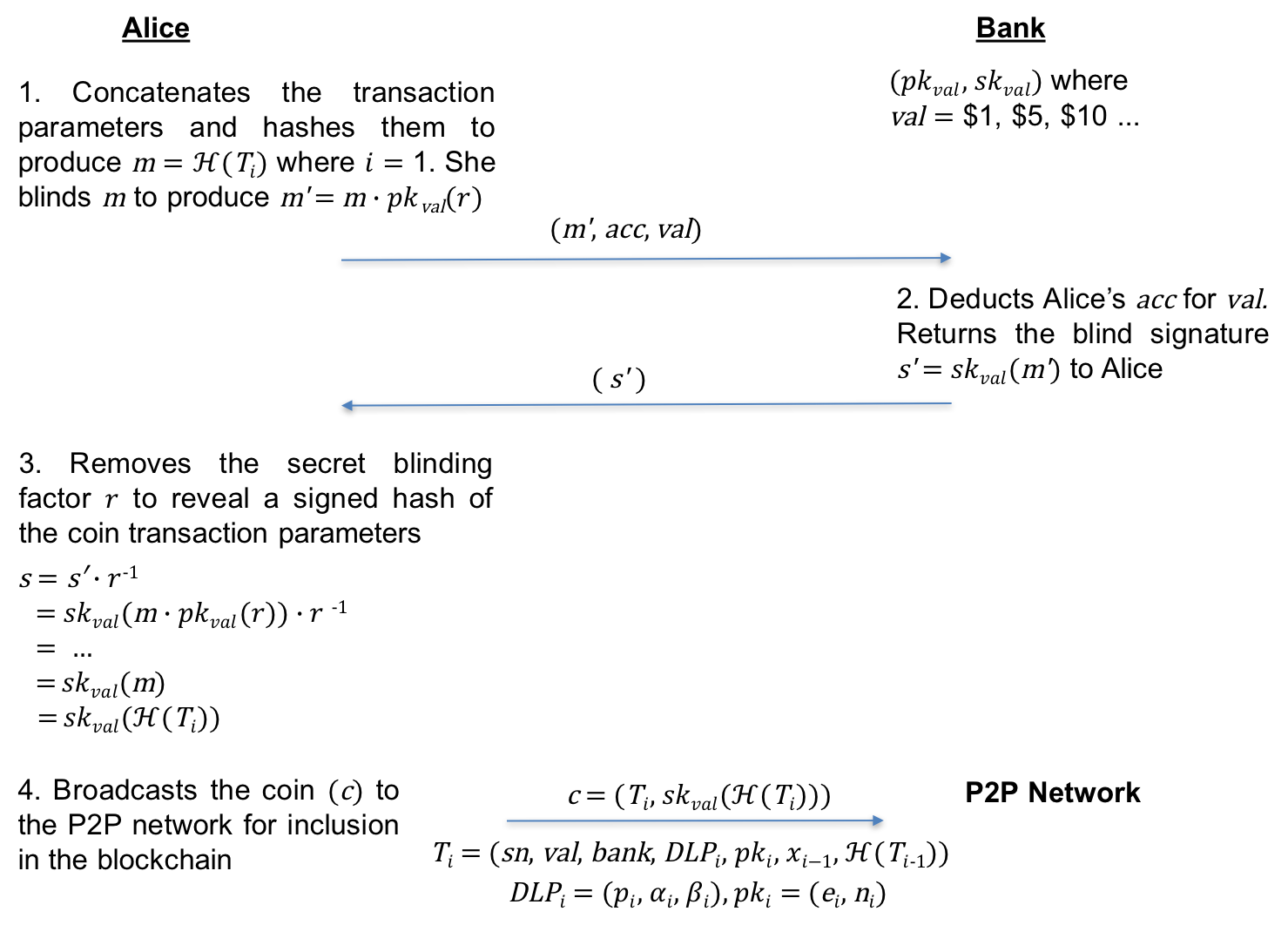}
\caption{Minting of Coins by a Bank}
\label{fig_minting}
\end{figure}

Figure \ref{fig_minting} shows the steps of how Alice gets her bank to blindly sign the first coin transaction $(T_i)$ where $i=1$ with its private signature key for a particular coin value $(sk_{val})$.

Alice constructs the coin's first transaction $T_i$ and then creates a hash of the transaction $m = H(T_i)$. She blinds this value with a random secret $(r)$ to construct $m'$, and sends this value to her bank along with her account number, and the value of the coin she requires.

Alice's bank has a number of signature keys for different denominational values (\$1, \$5, \$10, ...), and uses the private signature key $(sk_{val})$ corresponding to the $val$ parameter supplied by Alice during the first message exchange\footnote{All communications between Alice and her bank are conducted over a secure channel (e.g. SSL), and the bank has no record of the coin transactions that it just signed},\footnote{It is Alice's responsibility to ensure that she forms a correct coin transaction, as badly formed coins (e.g. where the $val$ parameter does not match the bank signature key $sk_{val}$) will not be accepted as a valid coin by the other nodes in the system}. The bank deducts the corresponding amount of fiat currency from Alice's account, and signs the coin's blinded first transaction $m'$ with its private signature key $(sk_{val})$. The bank then sends the blind signature $(s')$ in the second message exchange to Alice.

Alice removes her blinding factor $r$ from $s'$ to reveal the hash of the coin's first transaction signed by her bank. Alice now has the coin's first transaction signed by her bank without having revealed the contents of the transaction parameters to the bank. Only she knows the DLP secret $(x_i)$ which must be revealed by her in order to prove ownership of the coin. In addition, she alone knows the private ephemeral coin transfer key $(sk_i)$ which is used to transfer ownership of the coin to another user in the network. Alice broadcasts the newly minted coin on the P2P network to be included in the next transaction block of the global blockchain.

\subsection{Coin Transfer}
A coin in our scheme is a ``series of linked transactions" on the blockchain such that each subsequent coin transaction reveals the secret DLP parameter associated with the previous transaction which allows the payee to have the confidence that the payer is the owner of the coin. In addition, the transaction also consists of a signature by the previous owner assigning the coin to the new owner (i.e. payee) of the coin. This is achieved by the recipient by first downloading the latest version of the coin from the blockchain, using the coin's public parameters $(sn, val, bank)$ as an index.

The first transaction $(T_i)$ is always signed by the bank whose public key $(pk_{val})$ for value $(val)$\footnote{$val$ = \$1, \$5, \$10 ...} is known to all participants in the network. A user that asked the bank to mint the coin embeds their public coin transfer key $(pk_{i})$ into the transaction $T_i$. During subsequent coin transfers, the public key $(pk_{i+1})$ of the recipient of the coin is locked into the next transaction $(T_{i+1})$ by the owner when they create a blind signature, using their coin transfer private key $(sk_{i})$. We make use of a global ledger and the PoW algorithm to lock in valid coins into the blockchain, and prevent the double-spending of coins.

In order for Alice to transfer a coin to another user or spend coins at a merchant premises, she needs to fulfill the following two requirements:
\begin{itemize}
\item Alice must be able to \textit{prove ownership} of the coin that she trying to spend by revealing the DLP secret $(x_{i})$ for the last transaction $(T_{i})$ for the coin.
\begin{itemize}
\item Once the recipient (Bob) has verified the DLP $\alpha_i^{x_i} \equiv \beta_i~(mod~p_i)$, he generates a new transaction $(T_{i+1})$ to be signed with Alice's private coin transfer key $(sk_i)$. He does this by applying Alice's public coin transfer key $(pk_{i})$ to a random number $(r)$ to form a blinding factor $(pk_{i}(r))$\footnote{Alice's public coin transfer key $(pk_{i})$ can be found in the last coin instance on the blockchain}.
\item Bob blinds the hash of his new coin transaction $m = H((T_{i+1}))$ with the blinding factor to produce $m' = m * pk_i(r)$ and sends this to Alice.
\end{itemize}
\item Alice must \textit{transfer the coin} to Bob. Alice applies her private coin transfer key $(sk_i)$ to $m'$ to produce $s' = sk_i(m')$ and returns the blindly signed value.
\begin{itemize}
\item Bob removes $r$ to reveal the new coin transaction ($sk_i(H(T_{i+1}))$) signed with the private coin transfer key of Alice. 
\item Bob broadcasts the new version of the coin to the P2P network to be included in the next transaction block on the blockchain. Only when Bob sees the new version of the coin appearing in the blockchain does he complete the transaction.
\end{itemize}
\end{itemize}

\begin{figure}[h]
\centering
\includegraphics[width=3.5in]{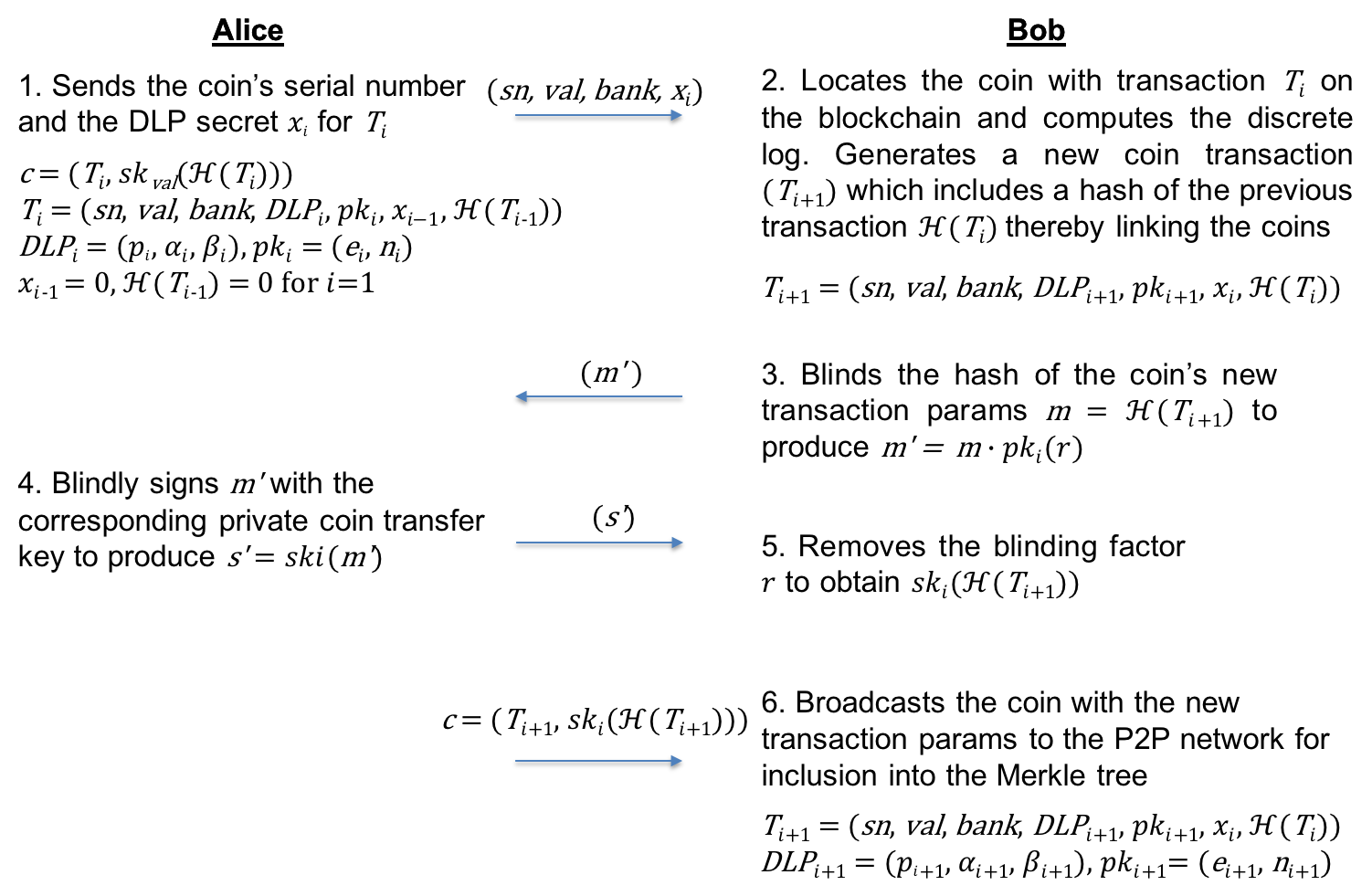}
\caption{Coin Transfer Protocol}
\label{fig_transfer}
\end{figure}

Figure \ref{fig_transfer} shows in detail the coin transfer steps between Alice and Bob. In the first message exchange Alice sends the coin's serial number $(sn)$, its value $(val)$, the financial institution that minted the coin $(bank)$, and the DLP secret $(x_i)$ for the last transaction $(T_i)$. Bob downloads the coin from the blockchain using the tuple $(sn, val, bank)$ as an index into the blockchain. He verifies that $x_{i}$ is the solution to the discrete logarithm problem $(DLP_i)$ of the last transaction $(T_i)$. As Alice did previously, Bob generates a new coin transaction which consists of a new set of DLP parameters $(\alpha_{i+1}, \beta_{i+1}, p_{i+1})$ and secret $x_{i+1}$ such that $\alpha_{i+1}^{x_{i+1}} \equiv \beta_{i+1}~(mod~p_{i+1})$, and a new ephemeral coin transfer key pair $(pk_{i+1}, sk_{i+1})$ which are only know to him such that the next transaction $(T_{i+1})$ is a follows:

\begin{center}
$T_i = (sn, val, bank, DLP_{i+1}, pk_{i+1}, x_{i}, H(T_{i}))$
\end{center}

The new transaction $(T_{i+1})$ consists of the serial number $(sn)$, the value of the coin $(val)$, the financial institution $(bank)$ that minted the coin, Bob's DLP public parameters $(DLP_{i+1})$, his public coin transfer key $(pk_{i+1})$, the DLP secret $(x_i)$ for the \textit{previous transaction} $(T_i)$, and a hash of the \textit{previous transaction} $H(T_i)$, thereby linking the two transactions together. He then blinds a hash of the new coin transaction $m = H(T_{i+1})$ by multiplying a blinding factor $pk_{i}(r)$, where $r$ is a random number and $pk_{i}$ is Alice's public coin transfer key, which can be obtained from the previous transaction $T_i$ to produce $m'$. Alice blindly signs $(m')$ with her private coin transfer key $(sk_i)$ and returns the blinded new transaction to Bob. Bob removes the blinding factor to reveal the new signed coin transaction $(sk_i(H(T_{i+1})))$ and broadcasts the new coin to the P2P network.

Once the coin is locked into the blockchain Bob knows that the transaction on the coin has been accepted by the network as being valid and he can now spend the coin, as only he knows the two secret quantities $(x_{i+1}, sk_{i+1})$ that allow him to prove ownership of the coin, and to transfer the coin legitimately to another user.

\subsection{Size of Coins}
We recall from section \ref{Overview}, that a coin in our system has the structure whereby the first three parameters of the coin remain constant, while the transaction list grows with each coin transfer:

\begin{center}$c = \langle T_i, sk_i(H(T_i))\rangle$ \end{center}
\begin{center} where: $T_i = (sn, val, bank, DLP_i, pk_i, x_{i-1}, H(T_{i-1}))$ \end{center}

\begin{itemize}

\item Params $(sn, val, bank)$ - 3 $*$ 256 bits

\item DLP public params $(p_i, \alpha_i, \beta_i)$ - 3 $*$ 1024 bits

\item Public key modulus $(n_i)$ - 2048 bits

\item Public key exponent and DLP secret $(e_i, x_{i-1})$ - 2 $*$ 256 bits

\item Hash of previous transaction $H(T_{i-1})$ - 256 bits
 
\item Blind signature $(sk_i(H(T_i)))$ - 2048 bits

\end{itemize}

We require 1088 bytes of storage for each instance of the coin on the blockchain. A user stores the tuple $(sn, x, sk)$ for each coin that they currently own - which amounts to the No. of coins $*$ 256+1024+2048 bits or 416 bytes of storage.

\section{Conclusion}
In this paper, we have presented an electronic cash scheme with unlimited transferability of coins. Our scheme prevents double-spending of coins within the system by making use of a blockchain to lock-in all valid transactions into a global ledger.

\end{document}